\documentclass{aa}  

\bibpunct{(}{)}{;}{a}{}{,} 

\usepackage{graphicx}
\usepackage[varg]{txfonts}
\begin{document}
   \title{On the assumption of Gaussianity for cosmological two-point statistics and parameter dependent covariance matrices.}


   \author{J. Carron
          \inst{1,2}
          }

   \institute{Institute for Astronomy, ETH Zurich, CH-8093 Zurich, Switzerland\\
         \and
           Institute for Astronomy, University of Hawaii, 2680 Woodlawn Dr., Honolulu, HI
96815, USA\\
\email{carron@ifa.hawaii.edu}
             }



\titlerunning{Gaussian likelihoods for two-point statistics}
 
  \abstract
  {
  In this brief paper we revisit the Fisher information content of cosmological power spectra or two-point functions of Gaussian fields in order to comment on the assumption of Gaussian estimators and the use of parameter-dependent covariance matrices for parameter inference in the context of precision cosmology. Even though the assumption of a Gaussian likelihood is motivated by the central limit theorem, we discuss that it leads to Fisher information content that violates the Cram\'er-Rao bound if used consistently, owing to independent but artificial information from the parameter-dependent covariance matrix. At any fixed multipole, this artificial term is shown to become dominant in the case of a large number of correlated fields. While the distribution of the estimators does indeed tend to a Gaussian with a large number of modes, it is shown, however, that its Fisher information content does not, in the sense that their covariance matrix never carries independent information content, precisely because of the non-Gaussian shape of the distribution.  In this light, we discuss the use of parameter-dependent covariance matrices with Gaussian likelihoods for parameter inference from two-point statistics. As a rule of thumb, Gaussian likelihoods should always be used with a covariance matrix fixed in parameter space, since only this guarantees that conservative information content is assigned to the observables, and at the same time prevents biases appearing.}
  
   {}
   {} 
   {}
   {}
   {}


   \maketitle
%
\newcommand{\beq}{\begin{equation}}
\newcommand{\enq}{\end{equation}}
\newcommand{\beqa}{\begin{eqnarray}}
\newcommand{\enqa}{\end{eqnarray}}

\newcommand{\veca}{\mathbf a}
\newcommand{\Tr}{\mathrm{Tr}}
\newcommand{\av}[1]{\left\langle #1 \right\rangle}
\newcommand{\lb}{\left [}
\newcommand{\rb}{\right ]}
\newcommand{\lp}{\left (}
\newcommand{\rp}{\right )}
\renewcommand{\max}{\mathrm{max}}
\renewcommand{\min}{\mathrm{min}}
\newcommand{\inv}{^{-1}}
\newcommand{\Fab}{F_{\alpha\beta}}
\newcommand{\bem}{\begin{bmatrix}}
\newcommand{\enm}{\end{bmatrix}}
\newcommand{\lmin}{\mathrm{l_{\min}}}
\newcommand{\lmax}{\mathrm{l_{\max}}}

\section{Introduction}
Starting from the second half of the nineties \citep{1996PhRvD..54.1332J,1996PhRvL..76.1007J,Tegmark97b,1997ApJ...480...22T}, calculating Fisher information matrices in order to understand the constraining power of an experiment quantitatively has become ubiquitous in cosmology, with its fundamental aspects now covered in cosmological textbooks (e.g. \cite{2003moco.book.....D,2008cmb..book.....D}, sections 11 and 6 respectively), or \citep{2009LNP...665...51H}. This is especially true for experiments aimed at measuring the power spectra of fields that are close to Gaussian fields, since in this case very handy analytical expressions can be obtained that can be applied in a variety of major cosmological subfields, such as the CMB, galaxy clustering, weak lensing, or their combination.
\newline
Nevertheless, even when applied to Gaussian variables, Fisher information matrices are not totally exempt from subtleties. We revisit the two different possible perspectives on the Fisher information content of such spectra. One starting point is often the assumption of Gaussian errors. We point out that this assumption of a multivariate Gaussian likelihood for power spectra estimators is not fully consistent with the purpose of understanding their information content, owing to a term violating the Cram\'er-Rao inequality, which we show is not necessarily small. Too much information is therefore assigned to the spectra under this assumption. We show that we can understand why this term is artificial precisely from the non Gaussian properties of the estimators, and discuss the reasons the usual formula, i.e. without this term, or with setting the covariance matrix to be parameter independent, still gives the correct amount of information. Our considerations apply indifferently to the spectra or the real space two-point correlation function. Namely, since the correlation function and the power spectrum are linearly related, the assumption of Gaussian power spectra is equivalent to assuming Gaussian two-point functions. We then comment on the role of the model parameter dependence of the covariance matrix within the Gaussian approximation, as studied in  \cite{2009A&A...502..721E} and \cite{2012ApJ...760...97L}.
\newline
\newline
In section \ref{sectionperspective} the two common approaches to the information content of spectra are discussed in detail in the case of a single field.  We clarify to what extent and why one is actually flawed, which is the source our comments on the use of Gaussian likelihoods and parameter-dependent covariances. In section \ref{severalfields} we then turn to a correlated family of fields, where the violation of the Cram\'er-Rao inequality is shown to become substantial. We summarize and conclude in section \ref{conclusions}.
\newline
\newline We recall first the specific form of the Fisher information matrix, defined for a probability density function $p$ as $\Fab = \av{\partial_\alpha \ln p\: \partial_\beta \ln p}, \alpha, \beta$ model parameters of interest, in the particular case of a multivariate Gaussian distribution with mean vector $\mu$ and covariance matrix $\Sigma$, \citep{1996ApJ...465...34V,1997ApJ...480...22T}
\beq \label{Fab} 
\Fab = \sum_{i,j}\frac{\partial \mu_i}{\partial \alpha}\Sigma^{-1}_{ij}\frac{\partial \mu_j}{\partial \beta} + \frac 12 \Tr \lb \Sigma^{-1}\frac{\partial \Sigma}{\partial \alpha}\Sigma^{-1}\frac{\partial \Sigma}{\partial \beta} \rb.
\enq
Remember that the Fisher information matrix has all the properties that a meaningful measure of information on parameters must have, most importantly for us here that any transformation of the data can only decrease its Fisher information matrix, regardless of the data distribution and parameter posterior. Thus, the Fisher information of the distribution of any estimator can only be lower than or equal to that of the data it is applied to.
\section{One field, gamma distribution}\label{sectionperspective}

Consider a zero mean isotropic homogeneous Gaussian random field, in euclidean space or on the sphere. It is well known that the Gaussianity of the field is equivalent to the fact that the Fourier or spherical harmonic coefficients are independent complex Gaussian variables, which are only constrained by the reality condition. Another equivalent description is that the real and imaginary parts of those coefficients form independent, real Gaussian variables. Such fields are described entirely by their spectrum, and so the extraction of the spectrum from the data with the help of an estimator is a fairly natural way to proceed for inference on parameters of interest. We place ourselves on the sphere, adopting the spherical harmonic notation for convenience. With the set of $a_{lm}$ the harmonic coefficients, the parameter-dependent spectrum $C_l$ is defined as
\beq
\av{a_{lm}a^*_{l'm'}} = \delta_{ll'}\delta_{mm'}C_l.
\enq
Standard, unbiased quadratic estimators can be written as a sum over the number of Gaussian modes available, as
\newline
\beq \label{estimator}
\hat C_l = \frac{1}{2l +1} \sum_{m = -l}^l |a_{lm}|^2.
\enq
We do not consider any source of observational noise, incomplete coverage, or any other such issue, because they are irrelevant for the points of our discussion.
\newline
\newline
At this point, there are two ways to approach the problem of evaluating the amount of information contained within the spectrum in the cosmological literature. The first - we call this approach the `field perspective' - calculates the information content of the field itself (equal to that of the set of $a_{lm}$'s), and then interprets this information as being the information within the spectrum. In this case, the information in the field is given by formula \eqref{Fab}, with zero mean vector and diagonal covariance matrix $C_l$:
\beq \label{Ffield}
F_{\alpha \beta} = \frac 12 \sum_{l = 0}^\infty (2l +1) \frac 1{C_l}\frac{\partial C_l}{\partial \alpha} \frac 1{C_l} \frac{\partial C_l}{\partial \beta},
\enq
where the factor $2l +1$ accounts for the number of independent Gaussian variables at a given multipole $l$.
The sum is in practice restricted to the multipole range that will actually be measured to obtain the information in the spectrum to be extracted. A very small sample of works using this approach are \cite{1997ApJ...480...22T}, \cite{2004PhRvD..70d3009H}, and \cite{2009ApJ...695..652B}. This approach is conceptually appealing, because it deals with the information content of the field itself and does not require either defining estimators or calculating their covariance. However, for the same reasons, it is only indirectly connected to data analysis since it is not yet specified precisely how this information content is to be extracted.
\newline
In the second approach - which we call the `estimator perspective' - an estimator $\hat C_l$ is defined first for each $C_l$ to be extracted, within some $\lmin$ and $\lmax$ (maybe with some bandwidth that we will not consider here), and its covariance matrix $\Sigma_{ll'} = \av{\hat C_l\hat C_{l'}} - \av{\hat C_l}\av{\hat C_{l'}} $ is calculated. Then it is argued that owing to the central limit theorem, the distribution of the estimator will be approximately Gaussian. In the case of the spectra of Gaussian fields, this is very well founded, at least for large $l$, since \eqref{estimator} is a large sum of well-behaved, identically distributed, independent variables. Then, under this assumption of Gaussianity, their information content is given by equation \eqref{Fab} with the mean vector this time the set of $C_l$ itself and (parameter-dependent) covariance matrix $\Sigma_{ll'}$,
\beq
\Fab = \sum_{l,l' = \lmin}^{\lmax} \frac{\partial C_l}{\partial \alpha}\Sigma^{-1}_{ll'}\frac{\partial  C_{l'}}{\partial \beta} + \frac 12 \Tr \lb \Sigma^{-1} \frac{\partial{\Sigma}}{\partial \alpha}\Sigma^{-1}\frac{\partial\Sigma}{\partial \beta}\rb.
\enq
It is well known that we have $\Sigma_{ll'} = \delta_{ll'}2C_l^2/(2l +1)$ for the estimator \eqref{estimator}. The Fisher information matrix, in the estimator perspective, thus reduces to
\beq \label{Festimator}
\begin{split}
 \Fab &=  \frac12 \sum_{l = \lmin}^\lmax (2l +1)\frac 1{C_l}\frac{\partial C_l}{\partial \alpha} \frac 1{C_l} \frac{\partial C_l}{\partial \beta} 
 \:+ \frac 12 \sum_{l = \lmin}^\lmax 4 \frac 1 {C_l} \frac{\partial C_l}{\partial \alpha} \frac 1{C_l} \frac{\partial C_l}{\partial \beta}.
 \end{split}
\enq
Clearly, the first term in the estimator perspective corresponds to that of the field perspective. However, the second term, which comes from the derivative of the covariance matrix, is new. That term is not enhanced by a $(2l +1)$ factor, and is therefore very subdominant at high $l$. It is either usually neglected, or the covariance matrix of the estimators is inconsistently taken to be parameter-independent, and in these cases the two approaches give the same results. Some expositions that explicitly use this perspective include \cite{Tegmark97b} and \cite{2003ApJ...598..720S,2007ApJ...665...14S}, where the additional term is neglected, or the approach in \citep[section 11.4.3]{2003moco.book.....D} where the covariance matrix is treated as parameter-independent. Works where this term plays the main role are \cite{2009A&A...502..721E} and \cite{2012ApJ...760...97L}, where the authors specifically study the impact of parameter dependent covariance matrices for parameter estimation using these Gaussian likelihoods.
\newline
\newline
Beyond the question of the quantitative relevance of this additional term, its very appearance is, however, very disturbing. Under this arguably reasonable Gaussian assumption, our estimator \eqref{estimator} is found to carry more information than the full field, even on the smallest scales.  This obviously violates the most fundamental property of Fisher information, i.e. that information can only be conserved at best when transforming the data (in this case reducing the field to its spectrum), a fact essentially equivalent to the celebrated Cram\'er-Rao inequality \citep{1997ApJ...480...22T}. Something must clearly have gone wrong in the assumption of a Gaussian likelihood for our spectra.
\newline
\newline
To understand what has happened, it is worth tracking the exact distribution and information content of the estimator \eqref{estimator}. Since they are independent at different $l$, we can work at a fixed $l$, and the total information content of these estimators will simply be the sum over $l$ of the information of the estimator at fixed $l$. Under our assumptions, the estimator is a sum of squares of $2l +1$ independent Gaussian variables, and its probability density function can be obtained with no difficulty. The exact distribution is the gamma probability density function with shape parameter $k$ and location parameter $\theta$ :
\beq \label{pgamma}
p(\hat C_l|\alpha,\beta) = \exp\lp -\hat C_l / \theta\rp\frac{ \hat C_l^{k-1}}{\theta^k \Gamma(k)},
\enq
with
\beq
\quad k = \frac 12 (2l +1),\quad \theta(\alpha,\beta) = \frac{2C_l}{2l +1},
\enq
and where $\Gamma$ is the gamma function. It is well known that the gamma distribution does indeed tend towards the Gaussian distribution for large $k$, with mean $\mu = k\theta = C_l$ and variance $\sigma^2 = k\theta^2 = 2C_l^2/(2l +1)$, as expected. However, its Fisher information content does not tend to that of the Gaussian. In our case, since only $\theta$ is parameter dependent, we find that the Fisher information in the estimator density function \eqref{pgamma} is
\beq
F_{\alpha\beta}^l = \frac{\partial \theta}{\partial \alpha}\frac{\partial  \theta}{\partial \beta}\av{\lp \frac{\partial \ln p (\hat C_l)}{\partial \theta}\rp^2}.
\enq
Since $\partial_\theta \ln p = (\hat C_l - k\theta)/\theta^2$, and $\partial_\alpha \theta = 2\theta \partial_\alpha C_l/C_l$, we obtain with straightforward algebra
\beq
\Fab^l = \frac 12(2l +1) \frac1 {C_l}\frac{\partial C_l}{\partial \alpha}\frac{1}{C_l}\frac{\partial C_l}{\partial \beta}.
\enq
Summing over $l$, we recover the first term of \eqref{Festimator}, but not the second. We have recovered the field perspective result \eqref{Ffield} at any $l$ without the Gaussian assumption but with the exact distribution. It turns out that even though the variance of the gamma distribution is parameter-dependent, it does not in fact contribute to the information. This can be seen as the following. Consider the information in the mean only of the estimator. From the Cram\'er-Rao inequality this must be less than the total information,
\beq
\frac 1 {\sigma^2}\frac{\partial \mu}{\partial \alpha}\frac{\partial \mu}{\partial \beta} \le \Fab^l.
\enq
Plugging in the values for the mean and variance actually leads to the result that the inequality is an equality,
so that the mean of the estimator captures all of its information.
\newline
\newline
In summary, the Gaussian approximation assumes  the mean and the variance of the estimator  are uncorrelated, such that both contribute to the information, while for the exact gamma, they are degenerate in such a way that the variance does not carry independent information.  Another way to see this, which we will use below when the exact form of the distribution is less convenient, comes from the fact that $\partial_\theta \ln p(\hat C_l)$ is a first-order polynomial in $\hat C_l$. It can be shown that the first $n$ moments capture all the information precisely when $\partial_\alpha \ln p$ is a polynomial of order $n$ \citep{2011ApJ...738...86C}.
\newline
\newline
That this function  $\partial_\theta \ln p(\hat C_l)$ is correctly reproduced by the Gaussian assumption with variance treated as fixed in parameter space has another interesting consequence that is relevant to cosmological parameter inference. Namely, performing parameter inference under that assumption does not shift maximum likelihood points $\partial_\theta \ln p(\hat C_l)= 0$, since this function is identical to that of the true likelihood \eqref{pgamma}. Therefore, no bias is introduced. This is no longer true if the variance is treated as parameter-dependent, where it is not difficult to see that the peak of the likelihood gets shifted by some amount decaying with $l$.

\section{Several fields}\label{severalfields}
It is instructive to see how these considerations generalize to a situation of a family of $n$ jointly zero mean Gaussian correlated fields, where the analysis proceeds through the extraction of the spectra and cross spectra.  In this case, the $C_l$ of the above discussion becomes an $n \times n$ (possibly complex) Hermitian matrix
\beq \label{multi}
\av{a_{lm}^ia^{j*}_{lm}} = \delta_{ll'}\delta_{mm'}C_l^{ij}, \quad C^\dagger = C.
\enq
From the hermiticity property, there are only $n(n+1)/2$ nonredundant spectra. Adequate estimators are defined by a straightforward generalization of equation \eqref{estimator},
\beq
\hat C^{ij}_l = \frac{1}{2l +1}\sum_{m = -l}^l   a^i_{lm}a^{j*}_{lm}.
 \enq
While the estimators are still independent for different $l$'s, the different components at a given $l$ are not. 
The information content of the set of $a_{lm}^i$ in the field perspective is still given by formula \eqref{Fab} for zero mean Gaussian variables. Explicitly, at a given $l$,
\beq \label{Fmean}
\Fab^l = \frac 12 (2l +1) \Tr \lb C_l^{-1} \frac{\partial C_l}{\partial\alpha}C_l^{-1}\frac{\partial C_l}{\partial\beta} \rb.
\enq
In the estimator perspective, assuming the estimators $\hat C_l^{ij}, i \le j$ are jointly Gaussian, we instead have
\beq \label{Fmean2}
\sum_{i<j,k<l = 1}^n\frac{\partial C_l^{ij}}{\partial \alpha}\Sigma^{-1}_{ij,kl}\frac{\partial C_l^{kl}}{\partial \beta}
+
\frac 12 \Tr \lb \Sigma^{-1} \frac{\partial{\Sigma}}{\partial \alpha}\Sigma^{-1}\frac{\partial\Sigma}{\partial \beta}\rb,
\enq
where the covariance matrix is
\beq
\begin{split}
\Sigma_{ij,kl}&= \av{\hat C_l^{ij}\hat C_l^{kl}} - C_l^{ij}C_l^{kl} 
=  \frac 1{2l +1}\lp C_l^{ik}C_l^{jl} + C_l^{il}C_l^{jk} \rp.
\end{split}
\enq
While it may not be immediately obvious this time, it has been noted (e.g \cite{2004PhRvD..70d3009H}) that the first term in \eqref{Fmean2} is rigorously equivalent to the expression from the field perspective \eqref{Fmean}. The estimator perspective under the assumption of a multivariate Gaussian distribution for $\hat C_l$ thus still violates the Cram\'er-Rao inequality due the presence of the second term. Since this term is not enhanced by a factor of $2l +1$ we expect it to be subdominant again. However, it is less true this time than in the one-dimensional setting: using the explicit form of the inverse covariance matrix,
\beq
\begin{split}
\Sigma^{-1}_{ij,kl} &=   \lp 2l +1 \rp\lp  C^{-1,ik}_l  C^{-1,jl}_l + C^{-1,il}_lC^{-1,jk}_l\rp \\
 &\quad\cdot \lp 1 - \frac12 \lp \delta_{ij} + \delta_{kl} \rp+ \frac 14 \delta_{ij}\delta_{kl} \rp,
\end{split}
\enq
one can derive, with some lengthy but straightforward algebra, the following expression for the violating term,
\beq\label{violating}
\begin{split}
 \frac 12 \Tr \lb \Sigma^{-1} \frac{\partial{\Sigma}}{\partial \alpha}\Sigma^{-1}\frac{\partial\Sigma}{\partial \beta}\rb 
 &= \frac 12 (n + 2) \Tr \lb C_l^{-1} \frac{\partial C_l}{\partial \alpha} C_l^{-1} \frac{\partial C_l}{\partial \beta}\rb \\
&\quad+ \frac12 \Tr \lb C^{-1}_l \frac{\partial C_l}{\partial \alpha}\rb\Tr \lb C^{-1}_l \frac{\partial C_l}{\partial \beta}\rb,
\end{split}
\enq
for any number $n$ of fields. If $n = 1$, we indeed recover \eqref{Festimator}. While the term is still subdominant at high $l$, the situation is still a bit less comfortable.  The number of fields is not necessarily very small in cosmologically relevant situations, such as tomographic joint shear and galaxy densities analysis in redshift slices, to which one may also add magnification, flexion fields, etc. Writing schematically $n = N_fN_{\textrm{bin}}$, where $N_{\textrm{bin}}$ is the number of bins and $N_f$ the number of fields per bin, we have, e.g., $N_f = 3$ for the galaxy density and the two shear fields, $N_f = 4$ including magnification, $N_f = 8$ adding hypothetically the four flexion fields, and so on.  Comparing \eqref{Fmean} and \eqref{violating}, and neglecting the second term in \eqref{violating}, we find that, at
\beq \label{lim}
l \sim \frac 12 N_fN_{\textrm{bin}},
\enq
the Cram\'er-Rao violating term is actually still the dominant one. This is still optimistic. Due to the product of two traces in the second term in \eqref{violating},  one can expect  roughly the same scaling with $n$ as the first term. Thus, the correct $l$ in \eqref{lim} may generically be closer to
\beq
l \sim  N_fN_{\textrm{bin}}.
\enq
From the discussion in section \ref{sectionperspective}, we can easily guess what went wrong. Consider the information content of the means of the estimators exclusively. This is given for any probability density function by weighting the derivatives of the means with the inverse covariance matrix, and is thus equal to the first term in \eqref{Fmean2}. Since already the means of the estimators do exhaust the information in the field, we can therefore already conclude that the total information content of the estimators must be equal to that of their means. In particular, the covariance matrix does not contribute to the information. As before, the second term in the estimator perspective is an artifact of the Gaussian assumption. It is interesting though to derive as above more explicitly why only the means carry information, from the shape of the joint probability density of the estimators. The remainder of this section sketches how this can be simply performed, leading to equation \eqref{form}.
\newline
\newline
For the sake of notation we restrict ourselves now to the case of two fields, $n = 2$ , but the following argumentation holds for any $n$.
The exact joint distribution for the three estimators $\hat C_l = (\hat C_l^{11},\hat C_l^{12},\hat C_l^{22})$, is given from the rules of probability theory as
\beq \label{pcl}
\begin{split}
&p(\hat C_l |\alpha,\beta) = \av{ \prod_{i\le j = 1}^2\delta^D \lp \hat C^{ij}_l - \frac 1 {2l +1}\sum_{ m = -l}^l a^{i}_{lm}a^{j*}_{lm} \rp }
\end{split}
\enq
where $\delta^D$ is the Dirac delta function. The average is over the joint probability density for the two sets of harmonic coefficients $a^1_{lm}$ and $a^2_{lm}$. Define the vector
\beq
\veca_l = (a^1_{l-l},\cdots,a^1_{ll},a^2_{l-l},\cdots,a^2_{ll}).
\enq
Since the $a_{lm}$ are zero mean Gaussian variables with correlations as given in \eqref{multi}, this probability density function is given by
\beq \label{palm}
\frac{1}{Z(\alpha,\beta)}\exp \lp-\frac 12 \veca_l^\dagger\cdot \mathbf C_l^{-1}\veca_l \rp,
\enq
with
\beq
\mathbf C_l = \bem C_l^{11}\cdot 1_{2l+1} &  C_l^{12}\cdot 1_{2l+1} \\  C_l^{21}\cdot 1_{2l+1} &  C_l^{11}\cdot 1_{2l+1}\enm,
\enq
where $1_{2l +1}$ is the unit matrix of size $2l +1$, and $Z(\alpha,\beta)$ is the normalization of the density for $\veca$, which does depend on the model parameters through the determinant of the $\bf C_l$ matrix. The inverse matrix $\bf C_l^{-1}$ has the same block structure, with entries those of $C^{-1}_l$. In the following we are not really interested in keeping track of the exact value of the components of this matrix, but only that they are dependent on the model parameters. With the understanding that $ C_l^{-1} =: D_l$, we have thus, due to the sparse structure of the $\mathbf C^{-1}_l$ matrix and the Dirac delta functions in \eqref{pcl},
\beq
\begin{split}
-\frac 12\veca_l^\dagger\cdot \mathbf C^{-1}_l\veca_l 
=-\frac12(2l +1)  \sum_{i,j = 1,2}D_l^{ij}\hat C_l^{ij} 
\end{split}
\enq
The presence of the Dirac delta functions means we can take the exponential \eqref{palm} out of the integral in \eqref{pcl}.  Writing the dependency of the different terms on $\hat C_l$ on the model parameters explicitly, we obtain the following form 
\beq \label{form}
\begin{split}
p(\hat C_l|\alpha,\beta) 
=  \frac {f(\hat C_l)} {Z(\alpha,\beta)} &\exp\lp -\frac12 (2l + 1) \sum_{i,j = 1,2}D_l^{ij}(\alpha,\beta)\hat C_l^{ij} \rp
\end{split}
\enq
which generalizes the gamma distribution, equation \eqref{pgamma}, in this multidimensional case . The factor $f(\hat C_l)$ is what is left from the integral \eqref{pcl} when the density for the set of $a_{lm}$ is taken out, i.e., the volume of the space spanned by the $a_{lm}$ that satisfies the constraints set by the Dirac delta function. It is thus a factor that depends on $\hat C_l$ but -important for us- not on the model parameters\footnote{The prefactors in \eqref{form} can be obtained in closed form, leading to the Wishart density function. See \citep[e.g.]{2008PhRvD..77j3013H}}. The point of the representation \eqref{form} is that it is immediately apparent that $\partial_\alpha \ln p(\hat C_l)$ is a polynomial first order in the components of $\hat C_l$. Second-order terms, which correspond to information within the covariance matrix, never appear, even though the exact density function becomes very close to Gaussian for large $l$.
It follows that the total Fisher information matrix is always equal to that of the mean, even if we did not derive the exact shape of the distribution.
\section{Summary and conclusions} \label{conclusions}
We discussed two common perspectives (the `field' and `estimator' perspectives) on the Fisher information content of cosmological power spectra and the reason the assumption of a Gaussian likelihood of the spectra estimators violates the Cram\'er-Rao inequality, by assigning the estimators more information than there is in the full underlying fields. Under the assumption of Gaussianity of the estimators, their means and covariance matrix are artificially rendered uncorrelated, creating an additional piece of information in their covariance, which we showed was inexistent by calculating the exact information content of the estimators' true probability density function. We showed that this violating term can become dominant in the limit of a large number of fields. Using Gaussian likelihoods consistently, i.e. with parameter-dependent covariance matrices as studied for example in \cite{2009A&A...502..721E} and \cite{2012ApJ...760...97L}, therefore assigns far too much information to the spectra in this regime, and should thus be avoided, as this allows tighter but artificial constraints on the parameters and can introduce biases as well.
A slight tightening of the constraints on parameters and tiny shifts in the parameters posterior maximum are indeed observed in these works. It is thus important to realize that these two effects do not reflect improvements over the method of treating the covariance matrix fixed, but should be considered spurious.
\newline
\newline
In the estimator perspective when deriving the Fisher information matrix, the piece of information coming from the covariance matrix is usually neglected. This paper clarified why it should not be present in the very first place and how the agreement between the field and estimator perspective can thus arguably be seen as a happy cancellation of two inconsistencies. It is interesting to note that the reason we still find the exact result in the estimator perspective without this wrong piece is that this expression is also the exact Fisher information content of the exact distribution of the estimators, which is strongly non-Gaussian for low $l$, the central limit theorem actually playing no role.
\newline
\newline
The other lesson we can take from this work is that when in doubt about the joint distribution of a set of estimators, a safe choice of information content is always that of their means, which only requires knowledge of their covariance. Provided the covariance matrix is correctly chosen, one is indeed certain, from the properties of Fisher information, to make  a conservative evaluation for any probability density function that does not rely on any further assumptions on its shape. Thus, leaving aside the question of the accuracy and consistency of the approximation itself, the use of a Gaussian likelihood with a parameter-independent covariance matrix, which has all of its information in the mean vector, remains a safe prescription in the sense that conservative information content is always assigned to the estimators.
\newline
\newline
Interestingly, the choice of a Gaussian distribution becomes motivated in this case not by the central limit theorem, but by conservative information content being assigned to the observables. From the Cram\'er-Rao bound, the constraints on the parameters using this assumption cannot be tighter than the ones allowed by the true distribution.  It is, however, essential in this respect that the covariance matrix is treated as fixed in parameter space. This holds true for any observables extracted from an arbitrary field distribution.\begin{acknowledgements} 
I wish to thank Cosimo Fedeli, Alexandre R\'efr\'egier, Marc Kamionkowski  and Istvan Szapudi, and to acknowledge the Swiss National Science Foundation, as well as NASA grants NNX12AF83G and NNX10AD53G for support.
\end{acknowledgements}

\bibliographystyle{aa}
 \bibliography{bib} 
\end{document}